# Semantic integration of UML class diagram with semantic validation on segments of mappings

Hicham Elasri[1], Elmustapha Elabbassi[2], Sekkaki Abderrahim[1], and Muhammad Fahad[3]

[1]Departement of Mathematics and Computer Science
University Hassan II, Ain Chock, Faculty of Sciences
Casablanca, Morocco

[2]Departement of Mathematics and Computer Science
University Mohamed V, Rabat, Faculty of Sciences
Rabat, Morocco

[3]Department of Computer Science,
University of Lyon, France

**Abstract.** Recently, attention has focused on the software development, specially by different teams that are geographically distant to support collaborative work. Management, description and modeling in such collaborative approach are through several tools and techniques based on UML models. It is now supported by a large number of tools. Most of these systems have the ability to compare different UML models, assist developers, designers and also provide operations for the merging and integration, to produce a coherent model. The contribution in this article is both to integrate a set of UML class diagrams using mappings that are result of alignment and assist designers and developers in the integration. In addition, we will present a detail integration of UML models with the validation of mappings between them. Such validation helps to achieve correct, consistent and coherent integrated model.

## 1   Introduction

The field of information systems engineering has always been a sector very claimant in techniques and new methods to improve both the quality of the products and the performance related to the process used for its development. In addition, applications are becoming increasingly complex and are covering wide range of fields. The increasing and diversifying role played by the Web and the Internet in the design and implementation of online applications (Semantic web and Cloud Computting) do amplify this situation.

This field, therefore, has evolved enormously since the advent of the world wide web and advancements in computer technologies. These developments have constantly provided reliable software and have been tailored to meet business needs, especially by reducing costs and delays. These developments have focused simultaneously on how to represent the targeted field by the software production, as well as conceptual, technological and methodological frameworks that facilitate and guide the process of software production.

In the past, information systems engineering programming activities were the first to be subject to this situation and to be experimented with this approach. Proposals based on models emerged later to address problems of engineering requirements, specification and modeling systems. Approaches based on Model-driven engineering (MDE) are part of this trend. They usually have similar concepts representing different objects. Their similarity and difference in terms of concepts, or their appearance necessitate a common specific interpretation of the models, their information and their general specifications.

Our problem is inscribed at the intersection of two scientific fields: the integration of models and the ontology alignment. The integration of models is a research problem that has been identified in the field of Model-Driven Integration (MDI) [28]. Our proposal aims to assist designers in the IS integration phase. Our novel integration process is guided by the background ontology (The Background ontology describes concepts and relations among concepts of the information system) to achieve semantic integration of models. This process allows the detection and resolution of semantics conflicts encountered in the process of integration of models also permit the validation of the mappings produced from matching of ontologies related to the candidate models.

We hypothesize that the conception of an Information System (IS) generally targets management business area and models represented in a high -level language such as UML. Moreover, semantic integration systems rely mostly on ontology alignment. We relied on the results of these studies to support the semantic integration process.

Our paper is organized as follow: our motivation is described in section 2. In the section 3 the ontologies and ontologies alignments is presented. In section 4 semantic integration of UML classes diagrams are described. In section 5 we present inconsistency, incompleteness and redundancy errors occur in the integration phase. Mappings validation for resolving in order to hide inconsistency presented in section 6. Our proposal of UML diagrams classes integration process is given in section 7. Finally, section 8 presents the conclusion and perspectives of our work.

## 2   Motivation

The automatic semantic integration of UML class diagrams derived from different sources involves detecting the semantic, syntactic and structural relationships. Semantic similarity measures play an important role for detecting different relationships in order to ensure alignment and merging of class diagrams. The majority of systems and alignment approaches address this problem by calculating the similarities between entities (concepts, roles, etc...) and produce candidate alignments based on the similarities obtained by comparing the entities one by one. Similarity-based methods using more sophisticated means for calculating spread and refine similarities speaking in the context of these entities. The alignment produces a set of mappings that can be used in the merging phase, but before that, a process of validating must be conducted in order to find incoherent and inconsistent mappings and keep only valid ones. This validation step is highly required to produce the correctness and consistency of final integrated model. Therefore, all steps of our semantic integration process are formalized by the mathematical approach in order to facilitate automation and validation of our semantic integration process.

## 3  Related work

Model integration is a crucial activity in Model Driven Engineering (MDE). Several approaches have been proposed to solve the problems of integration. These approaches have been proposed mainly in the field of heterogeneous databases integration. The work of models integration is focused primarily on syntactic and technical levels, very few studies, however, are concerned with the semantic aspect of the integration of conceptual models. Numerous works have adopted MDE techniques in order to define model integration as model transformations [31] [30].The AAM approach (Architecture Aspect Modeling) [32] proposes to use the directives of composition. However, The Theme approach [33] deals with the integration problem by combining a set of strategies for reconciliation upon a relationship kind of merge. Several types of reconciliation are offered (transformation, redefinition, etc.). The AMW approach [3] is based on model weaving and model transformation. It offers two resolution strategies: the first is automatic based on the use of heuristics to identify matches; another interactive, allows manual refining the correspondence links. The studies in the field of model differencing, which consist in a stage of integration; present syntactic differencing at either the concrete or the abstract syntax level. However, they not are able to represent the semantic differences between two versions of a class diagram [34].
In this paper, we propose a novel approach that addresses the model integration problem by means of graph segment, isomorphism of graphs segment and ontologies.

## 4  ONTOLOGIES AND ONTOLOGIES ALIGNMENTS

Ontologies are recently being developed for structuring knowledge and are defined as a collection of concepts and their interrelationships, which provide an abstract view of an application domain. According to Gruber, ontology is defined as an explicit formal specification of terms of a domain and relations among them [9]. Aligning ontologies consists in establishing semantic relations among concepts of various ontologies which describe the same or overlapping field of knowledge. Aligning ontologies represents a great interest in various application domains that manipulate heterogeneous overlapping knowledge, such as semantic web, communication in multi-Agent Systems, data Waterhouse, schemas/ ontologies integration [10], etc. Several works on the alignment of ontologies have emerged over recent years; most of them are based on an external resource that can be either a general ontology or domain ontology [11], [10]. In the following, we give an account of the concepts that we will use throughout the paper and in the metrics that we used for computing our alignments, mapping, graphs and sub-graph isomorphism.

**Ontology:** Ontology can be defined with different manners according to its type and use. In our case, we define ontology as a tuple$(C, R, T)$, where C is the set of concepts or OWL-classes, R is the set of relationships between concepts or OWL-properties, and T the set of relationships types.

**Concept:** A concept is an attribute vector V defined as $V = (T, At_1, \ldots, At_n, P_1, \ldots, P_m)$ where T is the concept label $At_i$ $(i = 1, \ldots, n)$ are the attributes that describe the concept. Finally, $P_j$ $(j = 1, \ldots, m)$ represent concept properties. They can be OWL datatype properties or object properties.

**Relationship:** A relationship is an attribute vector Vr defined as $Vr = (c_1, c_2, Tr)$ and $Vr \in R$, with $c_1$ and $c_2$ are two concepts $((c_1, c_2) \in C^2)$ and Tr is the type of the relationship between $c_1$ and $c_2$.

**Relationship type:** is the type of relationship, for example in UML, the relationship type are: inheritance, aggregation, composition.

**Similarity measure:** Similarity measure allows finding the semantic equivalence or similarities between entities. It is based on the concept terminology and properties. There are some approaches in literature for the classification of the Similarity measure [26]:

- **Syntactic indexes.** These indexes aim to detect the syntactical similarities among the various components of an ontology.
- **Semantic indexes.** These indexes aim to compare the ontologies from a semantic point of view using the WordNet taxonomy.
- **Structural indexes:** These indexes compare the ontologies from a structural point of view.

TABLE 1 : Example of classification of the similarity measure

| *Syntactic indexes* | *Semantic indexes* | *Structural indexes* |
|---|---|---|
| Editing Distance | Semantic Similarity | Attributes |
| Trigram Function | Granularity | Similarity Index for properties |
| Acronym | Synonym Index | Similarity Index for Entities |
| Fingerprint | Derived | |
| Abbreviation | Label | |
| Label | | |
| Attributes | | |

**Graph:** A graph $G = (V, E)$ comprising a V of vertices or nodes together with a set E of edges or lines.

**Subgraph:** Let $G(V, E)$ be a graph. A subgraph of G is a graph $G'(V', E')$ such as $V' \subset V$ and $E' \subset E$. For simplicity we write: $G' \subset G$

**Graphs Isomorphism:** Two graphs $G_1(V_1, E_1)$ and $G_2(V_2, E_2)$ are isomorphic if there is a bijection $f: V1 \to V2$ such as:

$$\forall a, b \in V_1, (a, b) \in E_1 \Leftrightarrow (f(a), f(b)) \in E_2$$

The function f is an isomorphism of graphs. For simplicity we write: $G_1 \cong G_2$.

In the case of ontologies (see definition above) relations are not of the same type; therefore the definition is as follows:

**Ontologies Isomorphism:** Two ontologies $O_1(C_1, R_1, T)$ and $O_2(C_2, R_2, T)$ are isomorphic if there is a bijection $f: V1 \to V2$ such as:

$$\forall a, b \in C_1 \text{ et } t \in T, (a, b, t) \in R_1 \Leftrightarrow (f(a), f(b), t) \in R_2$$

The similarity function of two graphs: Given two graphs $G_1(V_1, E_1)$ and $G_2(V_2, E_2)$, the similarity function is defined with a vertices based similarity measure, $b \in V_1$ is image of $a \in V_2$ if a is similar to b.

**Subgraphs isomorphism:** Given two graphs $G_1(V_1, E_1)$, $G_2(V_2, E_2)$, and given two sub-graphs respectively $G'_1(V'_1, E'_1)$ and $G'_2(V'_2, E'_2)$ of $G_1$ and $G_2$, It is said that $G'_1$ and $G'_2$ are isomorphic if they are isomorphic as graphs.

**Matching process:** A matching process can be seen as a function f which takes two ontologies O and O', a set of parameters P, and a set of oracles and resources R, and returns an alignment between O and O'.

**Mappings:** Given $O_1(C_1, R_1)$ and $O_2(C_2, R_2)$ two ontologies and α a measure of similarity between two concepts, a mapping is an element $(a, b)$ of $C_1 \times C_2$ such as $\alpha(a, b) > threshold$

**Set of mappings:** Given $\{O_i\}_{1<i<n}$ a family of ontologies. Set of mapping *Smappings* family is the set of all the feasible mapping between couples of ontology $(O_i, O_j)$ with $1 < i \neq j < n$

$$S_{mappings} = \{(a, b) \in C_i \times C_j, 1 < i \neq j < n \text{ et } \alpha(a, b) > threshold\}.$$

## 4.1 Transformation UML Class Diagrams into ontologies

A comparison between models and ontologies is given in [7]. The differences between the classes of the UML and OWL are studied in [8] and [9]. [14] Provided an analysis of approaches for transforming UML to ontologies. Transformations of UML models to ontologies can be grouped into three categories:

UML Extension [15] presents a UML extension that aims to improve the description (draped Agent Markup Language based) ontologies using UML. [16] Presents a graphical representation of OWL based on UML; it is extended by OWL annotation [17] presents a method for the automated determination of semantic relations between concepts of an ontology generated from conceptual models specified in UML.

- An approach based on XSLT: Gasevic [18] provided transformation rules of UML class diagrams into OWL ontologies. Transformations are performed by Extensible Stylesheet Language Transformation (XSLT) operating models in XML format. In addition, UML profile was used to model specific aspects of ontologies.
- Approaches based on Meta-Models: [15] described a meta-model for OWL based on Meta Object Facility (MOF) and a UML profile for modeling ontologies using UML. [19] is a preparation for the specification of the Ontology Definition Meta-Model (DOM).
- OWL meta-model as a UML profile [20] gave a transformation between UML and OWL ontologies based on Atlas Transformation Language (ATL). [21] used the MOF Script tool to perform the transformation from UML to owl2. However, they aimed at the validation of the Metamodels. They introduced several elements of UML in the ontology, which are necessary to achieve this goal, but it complicates the use of ontology. [23] proposed a methodology based on the driven engineering models (MDA) for the generation of ontologies from annotated UML business model. The authors presented a transformation of class diagrams in ontologies represented by owl2[13]. They specified the transformation in M2 using the QVT transformation language and metamodels and UML owl2.

Recently a promising new tool named umlTUowl proposed in European project (TwoUse Toolkit) for the transformation of class diagrams into an ontology [22]. umlTUowl has been developed to overcome the problems that arise during the transformation process of evaluated UML2OWL tools. It is not only optimized to transform UML class data models, as used by partners of CDL-Flex (Visual Paradigm for UML V7.2, 8.2; XMI 2.1, Microsoft Visio 2010 XMI 1.0) into OWL 2 DL ontologies. umlTUowl eases the integration of new transformation scripts, e.g. support for ArgoUML 0.32.2 XMI 2.1(freeware) has been already implemented. Its maxim is: don't try to provide an overall transformation solution for all vendors, because parsing of XMI is too fragile. Be prepared for variations and UML-model-vendor updates that affect the structure of the resulting XMI code by providing traceability of supported UML tools, as well as providing high testability, modifiability and extensibility. We will base on umlTUowl in the phase the transformation of class diagram into an otology of our integration process.

## 5   Semantic integration of uml classes diagrams

The semantic integration of different UML Class Diagram in the same information system goes through a process of detection and resolution of semantics conflicts that may exist among different models. We consider that each conflict is generated by a non-definition of a semantic relation (e.g.,: equivalence semantic relationship which may cause a conflict type naming). We based in this paper on ontologies alignment in order to align the ontologies related to UML Class Diagram, due to its ability of producing an ontology called Correspondant Ontology (CO) which includes the concepts and their semantic relationships derived from the multiple sources ontologies. This task is required and appropriate in the process of semantic integration.

That's why, we show the usefulness of CO and how it can be used either in an automatic process as an input of the integration phase as well as process assisted by the designers of information systems. This allows to deduct a set of actions (add, edit or delete a concept or relation) in order to achieve semantic integration of UML Class Diagram. The integration of UML Class Diagram aims to detect and resolve conflicts caused by the heterogeneity of UML Class Diagram. The goal is to produce a single unified model. We define the binary UML Class Diagram semantic integration based on semantic integration of ontologies related to the UML Class Diagram. We have proposed [1] and [2] an integration processes that reduces the problem of semantic integration of UML Class Diagram to a problem of ontologies alignment. Bezivin [3], defined the models integration as follow:

"*Takes two models MA, MB and a Correspondence Model CMAB between them as input and combines their elements into a new output model*".

We are based on this definition to define integration of UML Class Diagram: The integration of UML Class Diagram takes two models $BC_1$ and $BC_2$ and Correspondence Model $CM_{1,2}$ between them as input and combines their elements into a new output model $BC_{1,2}$:

Integration :

$$S_{models}^2 \times S_{C-models} \rightarrow S_{models}$$

$$(BC_1, BC_2, CM_{1,2}) \rightarrow BC_{1,2}$$

Integration is a binary integration, we rely on the latter to define the integration of a set of models, denoted $BC_1,...,BC_n$, takes a set of models: $BC_1,...,BC_n$ and correspondence model $CM_{1,n}$ between them as input and combines their elements into a new output component $BC_{1,n}$.

Integration:

$$Pf(S_{models}) \times S_{C-models} \rightarrow S_{models}$$

$$(\{BC_1,...,BC_n\}, CM_{1,n}) \rightarrow BC_{1,n}$$

$Pf(S_{models})$ is the set of finite subsets of $S_{models}$.

The semantic integration requires several preprocessing steps including transformation step of UML Class Diagram to ontologies and ontologies alignment step, resulting from transformation that fit into a phase called preIntegration that we can present it by function. The latter takes as input two UML Class Diagram: ($BC_1$ and $BC_2$ and a background ontology $DO_{1,2}$ to produce an Correspondence Ontology $CO_{1,2}$, which means:

**PreIntegration:**

$$S_{UML-D}^2 \times S_{D-Ontology} \rightarrow S_{C-Ontologies}$$

$$(BC_1, BC_2, DO_{1,2}) \rightarrow CO_{1,2}$$

We present the transformation step of UML Class Diagram to the ontology by function "Transformation" which takes as input a set of UML Class Diagram, $BC_1,...,BC_n$ to produce a set of ontologies $BCO_1,...,BCO_n$, which means:

**Transformation:**

$$Pf(S_{UML-D}) \longrightarrow P(S_{ontologies})$$

$$\{BC_1, ..., BC_n\} \longrightarrow \{BCO_1, ..., BCO_n\}$$

Pf(S) is the set of finite subsets of the set S.

We present ontologies alignment step of ontologies derived from UML Class Diagram by the function "alignment" that takes as input a set of ontologies $BCO_1,...,BCO_n$ and Background Ontology $DO_{1,n}$ and outputted correspondence ontology $CO_{1,n}$, which means:

**Alignment:**

$$Pf(S_{UML-D}) \times S_{D-Ontology} \rightarrow S_{C-Ontologies}$$

$$(\{BC_1, ..., BC_n\}, DO_{1,n}) \rightarrow CO_{1,n}$$

The semantic integration of models takes as input two UML Class Diagram: $BC_1$ and $BC_2$ and correspondence ontology resulting from "preintegration" $CO_{1,2}$ for produce a single UML Class Diagram Integrated $BC_{1,2}$, which means:

Integration:

$$S_{UML-D}^2 \times S_{C-Ontologies} \rightarrow S_{UML-D}$$

$$(BC_1, BC_2, CO_{1,2}) \rightarrow BC_{1,2}$$

Based on the binary integration we define semantic integration among UML Class Diagram:

**SemanticIntegration**: $Pf(S_{UML-D}) \times S_{C-Ontologies} \rightarrow S_{UML-D}$

$$(\{BC_1, ..., BC_n\}, CO_{1,n}) \rightarrow BC_{1,n}$$

### 5.1 Inconsistency, incompleteness and redundancy errors

In ontologies, there is a possibility of inconsistency, incompleteness and redundancy errors that may occur in a single ontology or arise in the merged global model due to conflicts and the semantic heterogeneities between source ontologies [27]. In this paper, we are more concerned to relate these errors to the merged model or integration phases.

Incompleteness in the merged model means that any important axiom or definition is missing in the merged model but that can be inferred from the source models. Redundancy in the merged model means that some information (axiom or definition) can be inferred more than once by different inferring mechanisms or modeled multiple times in the merged model. Inconsistency in the merged model means that any contradictory information can be deducted among the axioms and definitions of concepts between the merged model and the source models. Inconsistency is the most severe type of problem in the integration world that spoils the resultant merged model, therefore, in our contribution of semantic integration we are focusing on the semantic inconsistency errors.

Semantic Inconsistency in the merged ontology occurs when the merging system makes an incorrect class hierarchy by classifying a concept as a subclass of a concept to which it does not really (or partially) belong. This can happen due to conflicts; such as instantiation violation, property mismatches, subsumption violation, domain breach, or constraint dissatisfaction.

Another aspect which is crucial to avoid inconsistencies in the merging of heterogeneous ontologies is to analyze disjoint axioms between concepts in the source ontologies. On one side, their omissions create incompleteness in the global ontology. On the other side, they can create inconciseness (or redundancies) or may lead to semantic inconsistencies. Semantic Inconsistency such as common class between disjoint classes occurs when the merging system creates a class between two disjoint classes in the class hierarchy of concepts.

Detection of such errors in the pre-integration phase is vital and is done by the analysis of mappings so that their integration can result a conflict-free union of source models. Therefore, our aim is to integrate a semantic validation step in the pre-integration phase to get more reliable output.

## 6      Mappings validation for resolving inconsistency

Over the last decade researchers have debated the effectiveness of algorithms for the merging and the automatic integration of ontologies in order to circumvent the complexity of these tasks manual performing. Despite the great effort, the alignment and merging systems ontologies are still semi--automatic, which reduces the burden of creating and maintaining manual applications. These systems require human intervention to validate the alignment and merge ontologies. In addition, they use various aids, such as the common vocabulary, reference ontology, etc. [25] to detect mapping candidates. The validation process after the detection of initial mappings helps to find inconsistent mappings; which is usually done by domain experts and performed manually in most cases. During the validation phase, the domain expert is responsible for classifying mappings results from alignment based on a similarity measure in equivalence mapping and is-a mapping [24].
In this paper our goal is to validate the mappings resulted from the alignment of a set of ontologies related to UML Class Diagrams integration candidates. Only few studies have addressed the validation of ontology in several aspects business, semantic, structural and syntactic, among those who used the ontologies validation in the alignment process to detect some inconsistencies in the alignment relationships such as multiple correspondences and the following ones :

- ASMOV (Automated Semantic Mapping of Ontologies with Validation) [29] is an ontology alignment system designed to cooperate ontologies from heterogeneous data sources. ASMOV allows producing mappings between concepts and / or properties and / or instances of two ontologies. Then ASMOV uses a process of semantic validation for the candidate mappings, by checking the consistency of discovered correspondences with regard to the ontologies.
- RSMR (Reasoning Support for Mapping Revision) [5]: The semi-automatic technical review of the mapping aims to identify and repair invalid mappings. It proposes the use of formal methods based on the description logics to assist the user in the revision of invalid mappings automatically generated by alignment algorithms.
- MRVPR (Mapping validation by probabilistic reasoning)[4] : A probabilistic approach to perform mappings validation according to the semantics of involved ontologies.

The methodology of DKP -OM [25] proposed a unit for checking consistency of mapping and ensure that consistency of global merged ontology is generated by mappings, the unit are responsible for finding semantic inconsistency in the initial mapping when detectors discover any inconsistent mapping, they notify it to the ConsistencyChecker that warns the user about the inconsistent situations, which occur in global merged ontology by inconsistent following initial mapping. Hence, it reduces the human response by validating the merged ontology automatically.

The RSMR and MVPR techniques can be applied to detect invalid mapping after the ontology evolution. However, they require very formal ontologies expressed using logical languages. In addition, maintenance actions are limited to the removal of invalid mappings [6].

We present mappings based on correspondence ontology for the validation step generated from alignment step, that takes as input a set of mappings $CO_{1,n}$, the set of ontologies candidates for integration $\{BCO_1,...,BCO_n\}$ and the set of rules aims to validate and to detect inconsistence mappings $RV = \{rv_1, ..., rv_p\}$ and output a set validate mapping $COV_{1,n}$, which means:

**Validation**:

$$S_{C-Ontologies} \times Pf(S_{UML-D}) \times \{RV\} \to S_{C-Ontologies}$$

$$(CO_{1,n}, \{BCO_1, \dots, BCO_n\}, \{rv_1, \dots, rv_p\}) \to COV_{1,n}$$

$Pf(S_{UML-D})$ is the set of finite subsets of $S_{UML-D}$.

**Table 2: Some example for validation rules [25]**

## 7 UML diagrams classes integration process.

In this section we propose a solution of semantic integration of UML diagrams classes. The following are the prominent features of our solution:

| N° | Cause | Rules description |
|---|---|---|
| Rule 1 | Cycle between concepts | Let $\{A, B, C, \dots, L\} \in O_1$ and $\{M, N, O, \dots, Z\} \in O_2$ are concepts of ontologies Mapping $(A, M)$ and Mapping $(B, O)$ are inconsistent (create cycle in class hierarchy) when in $O_1$, $A \subset B$, and in $O_2$, $O \subset M$ |
| Rule 2 | Redundant concept subsumption | Mapping $(A, M)$, Mapping $(B, N)$ then Mapping $(C, L)$ create redundant subsumption when in $O_1$, $C \subset B \subset A$, and in $O_2$, $N \subset M$, $L \subset M$, but $L \not\subset N$. Or simply we can say, Mapping $(A, M)$, Mapping $(B, N)$ create redundant subsumption when in $O_1$, $B \subset A$ directly, and in $O_2$, $N \subset M$, but indirectly. |

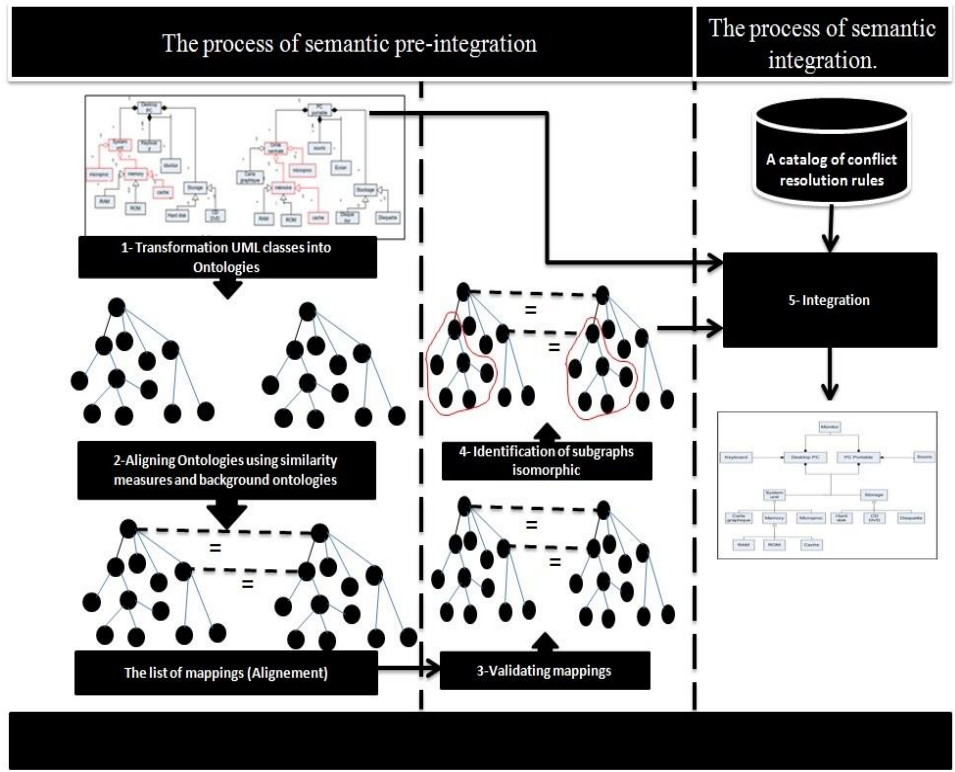

Figure 1 : semantic integration of uml diagrams classes

1. Transformation of UML class diagrams in ontologies

2. Use of a set of similarity measure to generate a set of mappings between ontologies on UML class diagrams

3. Validating mappings by set of validation rules

4. Identification of subgraphs isomorphic

    a. Construction of the set of mapping segments

    b. Construction of the set of equivalence classes of the relation bound by segments.

    c. Construction of the set subgraphs isomorphic.

5. Integration of UML class diagrams using subgraphs isomorphic and a catalog of conflict resolution rules

Produce a new BC resulting from the integration of UML diagrams classes sources.

Our proposal relies on the results of several research projects including those on the components transformation from a component modeling language into an ontology modeling language, and those related to the alignment ontologies [11], this solution consists of two complementary sub-processes:

- The process of semantic pre-integration.
- The process of semantic integration.

The objective of process perintegration is the production a set of semantic relation between concepts derived from the UML diagrams classes candidates for integration, represented by a Correspondence Ontology (CO). This process consists of a process description is provided in the following: The inputs of the integration process are:

- A set of UML diagrams classes selected by the designer in order to integrate them in the future Information system. We denote $\{BC_1,...,BC_n\}$ these UML diagrams classes.
- A Background ontology chosen by the designer according to the new IS.

The Background ontology describes concepts and relations among concepts of the IS. The Background ontology will thereafter be used to support the integration process.

The outputs obtained at the end of the Pre-integration process:

Correspondence ontology (Alignment): In the first step, IS designer can use this ontology to detect and resolve semantics conflicts in a semi-automatic process.

In the second step, the ontology could be reused in an automated process from the perspective of integrating UML diagrams classes while defining a set of integration rules derived from the correspondence of UML diagrams classes. It will later be used as the ontology support during the second phase: the integration process.

An correspondence ontology (Alignment) can be used as input in integration process

The pre-integration process comprises the following steps:

- Transformation the BC candidates for the integration into ontologies
- Aligning ontologies obtained based on the background ontology.
- Produce the correspondence ontology.

### 7.1 UML diagrams classes transformation into ontologies.

UML and OWL have similar concepts in many ways such as: classes, associations, properties, packages, types, generalization and instances. UML is used to model the dynamic behavior of a system. However, OWL does not allow this type of modeling. OWL is indeed able to infer navigating through relations between generalization and specialization classes, also individuals of a class based on the constraints imposed on the properties in the class definition, however, UML does not this feature [13]. A comparison between models and ontologies is given in [7].The differences between the classes of the UML and OWL are studied in [8] and [9].

## 7.2 Ontologies alignment.

Aligning ontologies is a crucial issue in the field of semantic integration, which aims to find semantic correspondences between a pair of elements of ontologies by identifying semantic relations.

The ontologies alignment use one or more similarity measures (syntactic, semantic and structural) to detect the set of mappings.

In the present paper our goal is to detect common parts between different ontologies on diagrams UML class. To better meet our objective, and to significantly increase the performance of integration algorithms following our approach, it is clear that the common parts that we should use are the largest parts. In the next section we define the maximum isomorphic sub-graphs and we propose a method of construction the set of these sub-graphs from the set of mappings result from the alignment of ontologies.

*a) Some specifications :*

- In this chapter, we will work with simple graphs but all properties and proposals are applicable in the case of ontologies.
- The work is done on a set of two graphs; therefore we assume that the constructing algorithm of the setof maximal isomorphic graphs will also work on the graphs two by two.
- We will say that $G_1$ is isomorphic to $G_2$ ($G_1 \cong G_2$) if the similarity function (see definition above) between the two graphs is an isomorphism.

*b) Maximum isomorphic sub-graphs*

Let $G_1(V_1, E_1)$, $G_2(V_2, E_2)$ be a graphs and $I(G_1, G_2) \coloneqq \{ (P_1, P_2) \setminus P_1 \subset G_1, P_2 \subset G_2, P_1 \cong P_2\}$. We called $I(G_1, G_2)$ the set of isomorphic sub-graphs of $G_1$ e $G_2$ (see Figure 1).

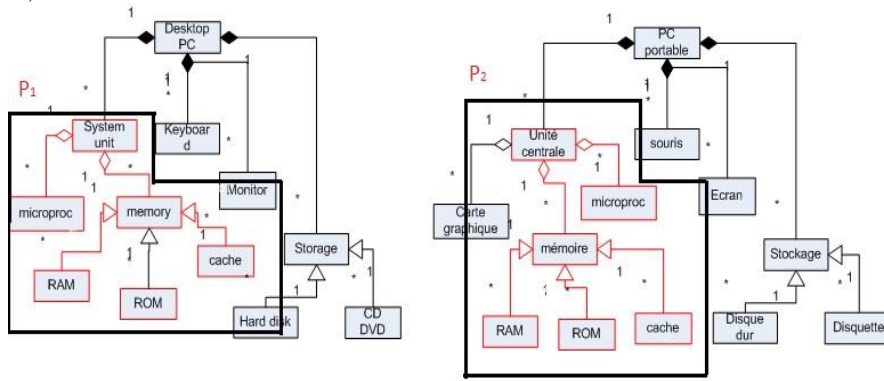

Figure 2 : Example of two maximum isomorphic subgraphs

Let $(P_1, P_2)$ and $(P_1', P_2')$ be two elements in $I(G_1, G_2)$ ; we say that $(P_1, P_2) \subset (P_1', P_2')$ if $P_1 \subset P_1'$ et $P_2 \subset P_2'$.

Inclusion ⊂ is an order relation in $I(G_1, G_2)$, which is finite, therefore let $Max(I)$ be the set of these maximal elements. For example in « Figure 1 » $(P_1, P_2)$ is a maximal element because we cannot find another element in $I(G_1, G_2)$ that contains it strictly (ie :$(P_1', P_2') \backslash (P_1, P_2) \subsetneq (P_1', P_2')$ et $P_1' \cong P_2'$).

$Max(I)$, by construction, contains the maximal isomorphic sub-graphs of $G_1$ and $G_2$, in the next section we will give a method that facilitates the construction of this set.

The relationship "Linking segment"

**Définition 1 (Segment):**

Let $G(V, E)$ be a graph, a segment S of G is a family of member of V, $\{a_i\}_{1 \leq i \leq n}$, such that for each $1 \leq i \leq n$, one of the two pairs, $(a_i, a_{i+1})$ and $(a_{i+1}, a_i)$, belongs to E.

In the case of ontologies, we must take into consideration the types of relationships, So the definition is as follows:

Let $O(C, R, T)$ be an ontology, a segment S of O is a family of member of C, $\{c_i\}_{1 \leq i \leq n}$, such that for each $1 \leq i \leq n$, one of the two vectors, $(c_i, c_{i+1}, t)$ and $(c_{i+1}, c_i, t)$, belongs to E (with $t \in R$).

**Notation 1**: we will write $S(a, b)$ for denote a segment with a and b as ends.

In « Figure 2 » we can see that the ordered set microproc, Système unit, memory, ROM} is a segment.

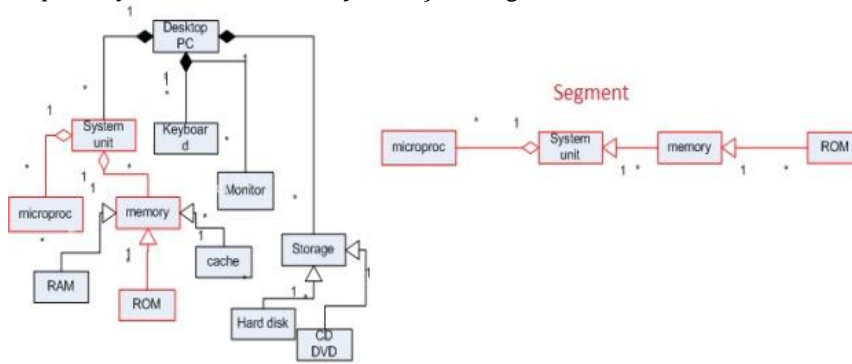

Figure 3 : Example of segment

The way by which we define the segment, shows us that we can see it as a sub-graph by adding to the set of $\{a_i\}_{1 \leq i \leq n}$ a set of relations which contains, for each $1 \leq i \leq n$, one of the two pairs, $(a_i, a_{i+1})$ or $(a_{i+1}, a_i)$, according to what is in the original graph (see « figure 2 »).

**Definition 2 (Isomorphism of segments)** : Let $G_1$, $G_2$ be two graphs and $S_1$, $S_2$ two segments of $G_1$ and $G_2$, We say that S1 and S2 are isomorphic if they are isomorphic as subgraphs (see « figure 3 »).

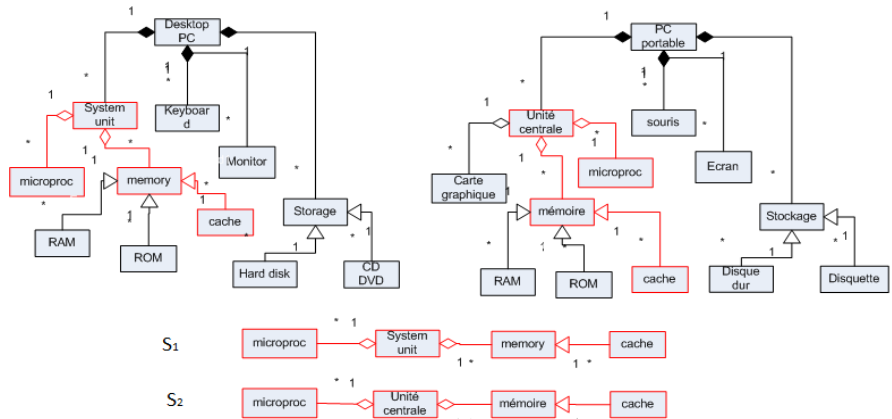

Figure 4 : Example of two isomorphics segments

**Definition 3 (Segment of mappings)** : Let $G_1(V_1, E_1)$, $G_2(V_2, E_2)$ be two graphs, and M the set of mapping of these two graphs, a segment of mappings S is a family of members of M, $\{(a_i, b_i)\}_{1 \le i \le n}$, Such that the two families $\{a_i\}_{1 \le i \le n}$ and $\{b_i\}_{1 \le i \le n}$ are two isomorphic segments of $G_1$ and $G_2$ (see « Figure 4 »).

**Notation 2:** We will write $S[(a, b), (c, d)]$ for denote a segment of mappings with (a,b) and (c, d) as ends.

Notation 3: Will write $S[(a, b), (c, d)]$ for denote the set of all possible segments of mappings in M.

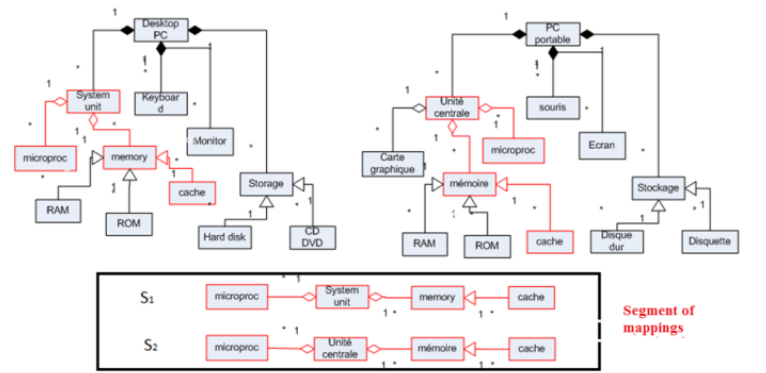

Figure 5: Example of mapping segment

**Definition 4 (The relation « bonding by segment»)** : Let M be the set of mapping for two graphs $G_1$ and $G_2$, We say that a mappings $(a_1, b_1)$ is bonded by segment with $(a_2, b_2)$, and we write $(a_1, b_1) <> (a_2, b_2)$, if there is a segment of mappings whose two ends are $(a_1, b_1)$ and $(a_2, b_2)$ (see Figure 5).

$$(a_1, b_1) <> (a_2, b_2) \equiv \exists\, S[(a1, b1), (a2, b2)] \in S(M)$$

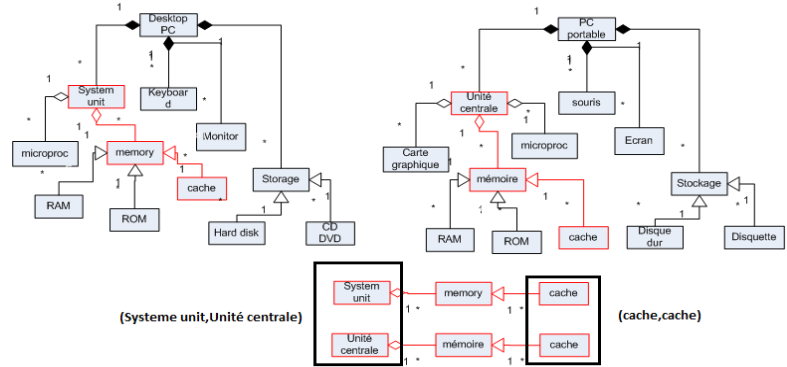

Figure 6 : Example relation bonding by segment between two mappings

**Proposition 1**

The relation « bonding by segment» is an equivalence relation.

Proof:

**Reflexive:** $\forall\ (a,b) \in M$ , $S[(a,b),(a,b)] \in S(M)$ therefore $(a,b) <> (a,b)$.

**symmetric:**

$(a1,b1) <> (a2,b2) \Leftrightarrow \exists\ S[(a1,b1),(a2,b2)] \in S(M)$

$\Leftrightarrow S[(a2,b2),(a1,b1)] \in S(M)$

$\Leftrightarrow (a2,b2) <> (a1,b1)$

**Transitive:**

Let $(a_1, b_1) <> (a_2, b_2)$ and $(a_2, b_2) <> (a_3, b_3)$ therefore $\exists S_1[(a_1, b_1),(a_2, b_2)] \in S(M)$ and $\exists S_2[(a_2, b_2),(a_3, b_3)] \in S(M)$. $S_1$ and $S_2$ can be written as families of mappings $\{(x\_i, y\_i)\}_{1 \leq i \leq n}$ and $\{(x\_i, y\_i)\}_{n \leq i \leq m}$, with $(x_1, y_1) = (a_1, b_1)$, $(x_n, y_n) = (a_2, b_2)$, $(x_m, y_m) = (a_3, b_3)$, since $(x_n, y_n)$ belongs to $S_1$ and to $S_2$, therefore the familie $\{(x\_i, y\_i)\}_{1 \leq i \leq m}$ constructed from two segments is a segment of mappings to, hence $(a_1, b_1) <> (a_3, b_3)$.

**Notation4:** Will write $C(M)$ for denote the set of equivalence classes of the relation $<>$ and $\overline{(a,b)}$ for denote equivalence class of mapping $(a, b)$.

**Proposition 2:**

The set of equivalence classes of the relation $<>$, $C(M)$, is identical (up to a bijection) to the set of the maximal isomorphic sub-graphs of $G_1$ and $G_2$, $Max(I)$.

**Proof :**

Let the function:

$$F : C(M) \rightarrow Max(I)$$
$$\overline{(a,b)} \rightarrow (P_1, P_2)$$

Such that $\overline{(a,b)} \subset V_1 \times V_2$ with $V_1$ is the set of the vertices of $P_1$ and $V_2$ the set of the vertices of $P_2$.

Show that F is well-defined:

Let $\overline{(a, b)}$ an element of $C(M)$. $\overline{(a, b)}$ contains all the elements bonded by segment with $(a, b)$, and by definition each segment of mapping is constructed by two isomorphic segments, so we can build two isomorphic subgraphs of $G_1$ and $G_2$, $P'_1(V'_1, E'_1)$ et $P'_2(V'_2, E'_2)$ such that $V'_1 = \{x : \exists(x, y) \in \overline{(a, b)}\}$ and $V'_2 = \{y : \exists(x, y) \in \overline{(a, b)}\}$.

Now we'll show that $(P'_1, P'_2)$ is maximal in $I(G_1, G_2)$ : Let $P_1(V_1, E_1)$ et $P_2(V_2, E_2)$ such that $(P_1, P_2) \in I(G_1, G_2)$ and $(P'_1, P'_2) \subset (P_1, P_2)$ and let x similar to y with $(x, y) \in V_1 \times V_2$. As $a \in V'_1 \subset V_1$, $b \in V'_2 \subset V_2$ and $P_1$ and $P_2$ are isomorphic, then $(x, y)$ and $(a, b)$ are linked by segment, then $(x, y)) \in \overline{(a, b)}$, then $x \in V'_1$ et $y \in V'_2$ and then $(P_1, P_2) \subset (P'_1, P'_2)$ therefore $(P_1, P_2) = (P'_1, P'_2)$ hence $(P'_1, P'_2)$ is maximal and $F(\overline{(a, b)}) = (P'_1, P'_2)$.

Show that F is bijective :

**Surjectivity :** Let $(P_1, P_2) \in \text{Max}(I)$, then $P_1(V_1, E_1) \cong P_2(V_2, E_2)$ therefore there is an element $(a, b)$ of the set of mapping M, such that $(a, b) \in V_1 \times V_2$, and by the same method above can be shown that $F(\overline{(a, b)}) = (P_1, P_2)$.

**Injectivity :** Suppose that $F(\overline{(a, b)}) = F(\overline{(c, d)}) = (P_1, P_2)$, then $(a,c) \in V_1^2$ the set of vertices of $P_1$, therefore that is a segment $S_1 = \{a_i\}_{1 \leq i \leq n}$ of $P_1$ with $a_1 = a$ and $a_n = c$. $(P_1, P_2) \in \text{Max}(I)$ then isomorphic then that is a segment $S_2 = \{b_i\}_{1 \leq i \leq n}$ of $P_2$ isomorphic to $S_1$, which means that $(a, b_1)$ and $(c, b_n)$ belongs to the set of mappings M (a is similar to $b_1$ and c is similar to $b_n$), so as $(a, b)$ and $(c, d)$ belongs to M and also a graph can not contain two similar vertices, then $b_1 = b$ and $b_n = d$, therefore it is possible to form a segment of mapping by $S_1$ and $S_2$ whose two ends are $(a, b)$ and $(c, d)$, hence $\overline{(a, b)} = \overline{(c, d)}$.

**Remark:** the previous proposition gives a very interesting result in this study, in the following we will define a function that share the same principle to give birth to the algorithm of construction of the set Max (I).

Identification of subgraphs isomorphic

Let $G_1(V_1, E_1)$, $G_2(V_2, E_2)$ and M the set of mapping of this two graphs.

**Definition 5:** The "Equivalence Class Function" is a function that allowed us to browse the mappings linked by segment with a fixed member of the set of mapping M, which gives as a result its equivalence class, this function is of M into P(M) the set of all subset of M:

$$\text{ECF} : M \to P(M)$$

$$(a, b) \to A \in P(M)$$

With $A = \{(a', b') \in M\ /\ \{(a, b), (a', b')\} \in S(M)\}$.

Now let $(a, b) \in M$, By recursivity we can define the sequence $(A_n)_{n \in \mathbb{N}}$ in $P(M)$ such that :

$$\begin{cases} A_0 = \{(a, b)\} \\ A_n = \text{ECF}(A_{n-1}) \setminus \bigcup_{0 \leq i \leq n} A_i \end{cases}$$

By construction the $A_n$ are disjoint and M is finite, so from a certain rank R the $A_n$ will be empty, this can be used as a test for stoping the recursive function.

We will call R the rank of $(a, b)$ and we denote it by $rg(a, b)$.

**Proposition 3 :**

Let $(a, b) \in M$ and $(A_n)_{n \in \mathbb{N}}$ the sequence defined by $(a, b)$ as above, then $\bigcup_{n \in \mathbb{N}} A_n = \overline{(a, b)}$.

**Proof :**

Let $(a', b') \in \bigcup_{n \in \mathbb{N}} A_n$ then $(a', b') \in A_{n_0}$ for some $n_0 \in \mathbb{N}$, by construction of $A_n$ we can say that there is an element of $A_{n_0-1}$ bonded by segment with $(a', b')$, and so on until we reach $A_0 = \{(a, b)\}$ then $(a', b') <> (a, b)$ hence $(a', b') \in \overline{(a, b)}$ and we conclude that $\bigcup_{n \in \mathbb{N}} A_n \subset \overline{(a, b)}$.

Let Now $(a', b') \in \overline{(a, b)}$ then $(a', b') <> (a, b)$ then there is a segment of mappings $\{(a_i, b_i)\}_{1 \leq i \leq n}$ such that $(a_1, b_1) = (a, b)$ and $(a_n, b_n) = (a', b')$ then $(a', b') \in A_n$ hence $(a', b') \in \bigcup_{n \in \mathbb{N}} A_n$ and we conclude that $\overline{(a, b)} \subset \bigcup_{n \in \mathbb{N}} A_n$.

**Remarks:**

Knowledge of ranks of elements of M in advance (where an approximation), can help us in the choice of the inputs of algorithm, because we can gain a lot in terms of performance by choosing the elements of smallest ranks.

The function ECF facilitates algorithm construction of $C(M)$ that is identical to the set of maximal isomorphic subgraphs Max (I) (proposition 3).

Semantic integration process

The alignment process BC candidates for the integration outputs a Correspondence Ontology (CO) between the concepts of BC. The Correspondence Ontology eventually externalize resource or support ontology for the integration process, which can help developers to achieve their information system design tasks or integrate BC automatically using a set of rules for resolving conflicts that operate the semantic relationships in CO. The Correspondence Ontology is used as a new ontology support in future iterations in the integration phase there by increasing the efficiency of the process.

A set of UML diagrams classes, denoted as BC1, BC2, ... BCn selected by the designer for their integration in the future information system.

A correspondence ontology between the concepts of BC is the result from the pre-integration phase. The semantic integration often requires to find the correspondences between the entities: components, classes, attributes, services. It is in this context that we proposed to create an ontology that includes correspondence of concepts candidates for integration and relationships starting and semantic relationships detected in the pre-integration.

A catalog of conflict resolution rules, which includes a set of resolution rules (e.g., for the conflict type homonomie resolution rule is the re-naming by different names, if the concepts are synonymous must remove one of the two) by default which operate according to the types of conflicts.

We consider that every conflict is generated by a non-definition of a semantic relation (e.g., synonymy semantic relationship which may cause a conflict type naming).

# 8   Example

The purpose of this example is to explain the steps respectful of our approach. We will work on two simple graphs G_1 and G_2 (see Figure 7).

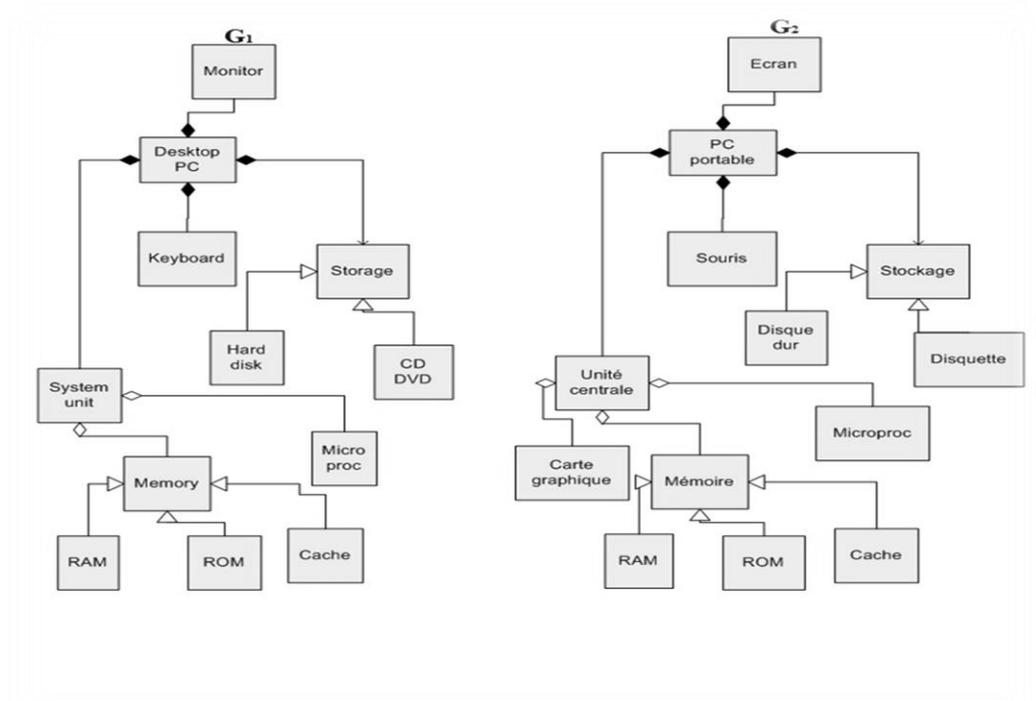

Figure 7 : Graphs G_1 and G_2

The first step is to transform UML class diagrams into ontologies. So the result is O_1 (C_1,R_1,T_1) and O_2 (C_2,R_2,T_2), with
C_1= {Desktop PC,Keyboard, System unit,…}
R_1= {(Desktop PC,Keyboard, Composition),…}
T_1= {Inheritance, Aggregation, Composition}
C_2= {PC portable, Souris,Unité centrale,…}
R_2= {(Stockag, Disque dur, Inheritance),…}
T_2= {Inheritance, Aggregation, Composition}
Use of a similarity measure to generate a set of mappings between ontologies on UML class diagrams M
M= {(Monitor,Ecran),(Storage,Stockage),
(Hard disk,Disque dur),(RAM,RAM),…}

After validating mappings by set of validation rules, we will use the ECF algorithm (see VII.B.d), based on segments mapping, to identify the equivalences class's. In "Figure 8" we have some example for a mapping segments:

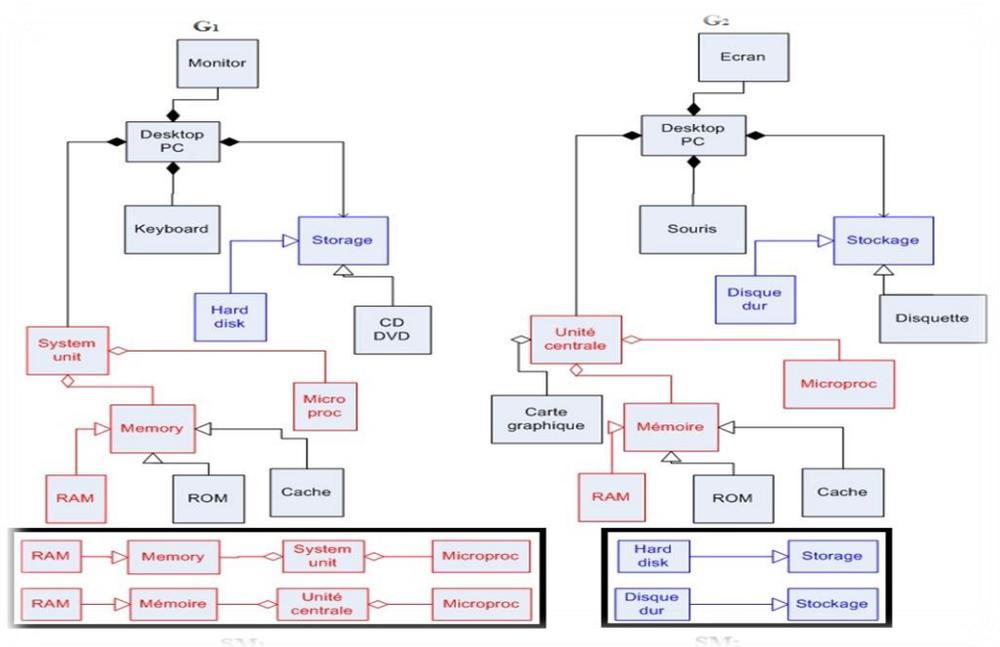

Figure 8: Mappings Segments

For example the following mappings are bonded by segment (RAM,RAM) and (Microproc, Microproc) (Memory, Mémoire) and (Microproc, Microproc) (ROM,ROM) and (System unit, Unité centrale) (Hard disk, Disque dur) (Storage, Stockage)
The result of this step is the set of equivalences class's C(M)
C(M)={{(Monitor,Ecran)},{(Storage,Stockage),
(Hard disque,Disque dur)},{(RAM,RAM),(ROM,ROM),(System unit,Unité centrale),(Cache,Cache),(Memory,Mémoire),(Microproc,Microproc)}}.
By "proposition 3" C(M) is identical to the set of maximals isomorphic subgraphs I(M) presented in "Figure 9".

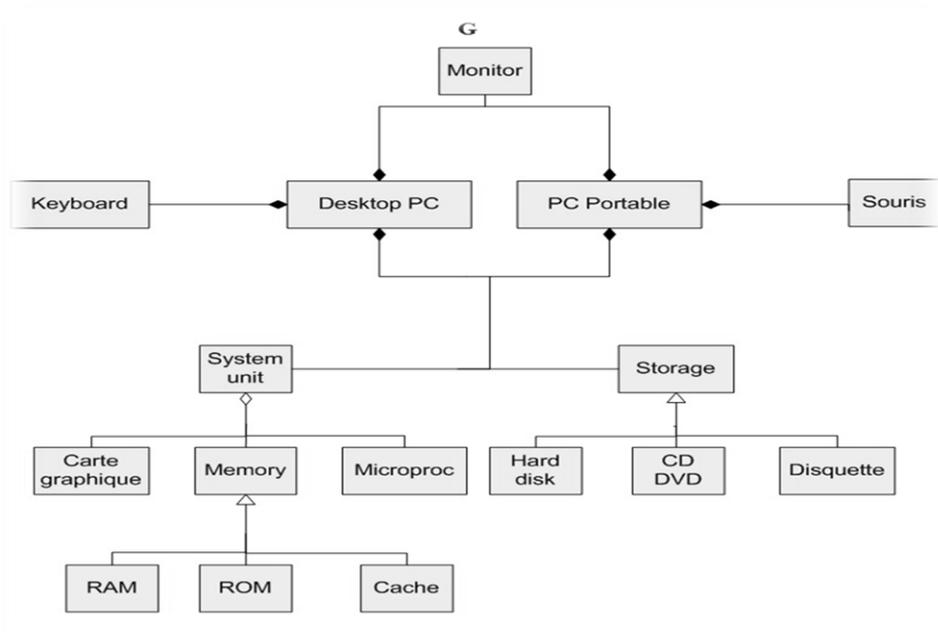

Figure 9: maximals isomorphic subgraphs

Now, using the set of maximals isomorphic subgraphs, we can construct the graph G, result of integration of G_1 and G_2 (see "Figure 10") .

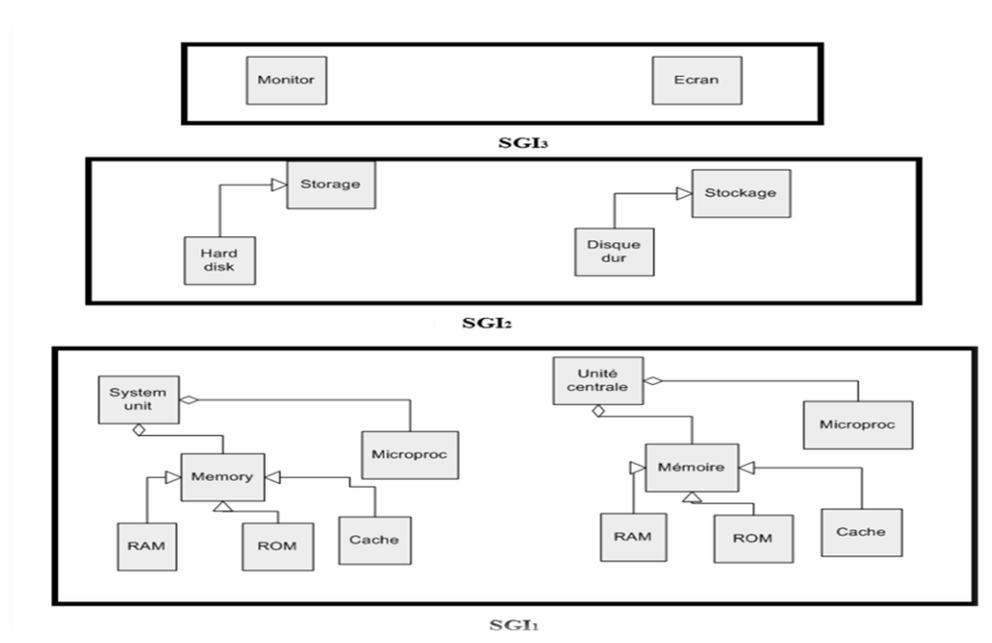

Figure 10 : The diagram G result of integration of G_1 and G_2.

## 9 Conclusions

In this paper, we contributed semantic integration of UML class diagram with the semantic validation based on the segments of mappings. We proposed the naming semantics conflict resolution of UML classes diagrams in the conceptual phases of analysis and design. Our novel solution for the semantic integration of UML class diagrams is based on the comparison of the source UML class diagrams in the pre-integration phase through concepts measurement. We contributed a set of validation rules for the detection of consistent and correct mapping derived from the alignment of ontologies related to the UML class diagrams. Our methodology is composed of segments measurement that is formalized using mathematical approach. We hope that our solution is a significant milestone towards the semantic integration of UML class diagrams to enable interoperability among multi-vendor engineering information systems.

Our ongoing research is to analyze semantic heterogeneities among UML class diagrams and integrate other semantic errors (such as incompleteness and redundancies) in the semantic integration phase to achieve the global model with higher level of quality.